\documentclass[letterpaper]{JHEP3}
\usepackage{amssymb,amsfonts}
\usepackage{cite}
\usepackage{epsfig}

\newcommand{\R}{{\bf R}}

\newcommand{\CA}{{\cal A}}
\newcommand{\CD}{{\cal D}}

\newcommand{\CL}{{\cal L}}

\newcommand{\CR}{{\cal R}}

\newcommand{\p}{\partial}

\renewcommand{\tilde}[1]{\widetilde{#1}}
\newcommand{\be}{\begin{equation}}
\newcommand{\ee}{\end{equation}}
\newcommand{\bea}{\begin{eqnarray}}
\newcommand{\eea}{\end{eqnarray}}
\usepackage{graphicx}

\title{Lifshitz Gravity for Lifshitz Holography}
\author{Tom Griffin${}^{a,b}$, Petr Ho\v{r}ava${}^{a,b}$ 
and Charles M. Melby-Thompson${}^c$
\\
${}^a$Berkeley Center for Theoretical Physics and Department of Physics\\
University of California, Berkeley, CA 94720-7300, USA\medskip\\
${}^b$Physics Division, Lawrence Berkeley National Laboratory\\ 
Berkeley, CA 94720-8162, USA\medskip\\
${}^c$Kavli Institute for the Physics and Mathematics of the Universe\\
University of Tokyo, Kashiwa 277-8583, Japan}
\abstract{
We argue that Ho\v{r}ava-Lifshitz (HL) gravity provides the minimal 
holographic dual for Lifshitz-type field theories with anisotropic 
scaling and dynamical exponent $z$.  First we show that Lifshitz spacetimes 
are vacuum solutions of HL gravity, without need for additional matter.  
Then we perform holographic renormalization of HL gravity, and show how it 
reproduces the full structure of the $z=2$ anisotropic Weyl anomaly in dual 
field theories in $2+1$ dimensions, while its minimal relativistic gravity 
counterpart yields only one of two independent central charges in the 
anomaly.}
\begin{document}
The concepts of scaling and the renormalization group have played a 
central role in organizing our understanding of quantum field theory (QFT) and 
statistical systems for half a century.  Here we will focus on systems in 
$D+1$ spacetime dimensions which exhibit scaling anisotropic between time 
and space, 
\be
\label{scaling}
t\to b^zt,\qquad x^i\to b x^i,\quad i=1,\ldots D,
\ee
with the degree of anisotropy measured by the dynamical exponent $z$.  
Systems with such Lifshitz scaling appear frequently in quantum and 
statistical field theory of condensed matter systems \cite{lubensky}, 
especially in the context of Lifshitz multicritical points, and in 
nonequilibrium statistical mechanics.  More recently, in a seemingly unrelated 
development, anisotropic Lifshitz-type 
scaling (\ref{scaling}) has played a central role in the new approach to 
quantum gravity initiated in \cite{mqc,lif} and commonly referred to as 
Ho\v{r}ava-Lifshitz (HL) gravity.  

Since AdS/CFT correspondence taught us that many relativistic QFTs have 
relativistic gravity duals, it seems natural to expect that the two disparate 
applications of Lifshitz scaling -- nonrelativistic QFT on one hand and HL 
gravity on the other -- should similarly be related by a holographic duality.  
The background geometry that captures the spacetime symmetries of QFTs with 
Lifshitz scaling (\ref{scaling}) is easy to find; it is given by the Lifshitz 
spacetime \cite{klm} in $D+2$ dimensions, 
\be
\label{lifmet}
ds^2=-\left(\frac{r}{\ell}\right)^{2z}dt^2+\left(\frac{r}{\ell}\right)^2 dx^idx^i
+\left(\frac{\ell}{r}\right)^2{dr^2}.
\ee
(From now on, we will set its radius of curvature $\ell=1$ for convenience.)
The holographic gravity duals of Lifshitz-type QFTs should therefore have 
(\ref{lifmet}) as their solution.  

Until now, the overwhelming share of work on Lifshitz holography (starting 
with \cite{klm}) does not use HL gravity -- it uses relativistic bulk gravity 
coupled to matter instead.  In the relativistic case, the coupling to matter 
is necessary, as the Lifshitz spacetime with $z\neq 1$ does not solve the 
Einstein equations in the vacuum.  Here we stress that another natural 
option is available: Instead of adding {\it ad hoc\/} matter to Einstein 
gravity so that (\ref{lifmet}) becomes a solution, one can modify gravity 
itself.  

Perhaps the most popular relativistic model for Lifshitz holography, proposed 
in \cite{taylor}, consists of Einstein gravity (described by the bulk metric 
$G_{\mu\nu}$, in coordinates $y^\mu=(t,x^i,r)$, with $\mu=0,\ldots D+1$), 
coupled to a massive vector $A_\mu$:
\be
\label{rela}
S_{\rm rel}=\frac{1}{16\pi G_{\rm N}}\int dt\,d^Dx\,dr\sqrt{-G}\left\{\CR
-2\Lambda-\frac{1}{4}F_{\mu\nu}F^{\mu\nu}
-\frac{1}{2}m^2 A_\mu A^\mu\right\}+{\rm surface\ terms}.
\ee
The Lifshitz geometry (\ref{lifmet}) is a solution for an appropriate 
condensate of $A_0$ and an appropriate choice of $\Lambda$ and $m$.    

In this paper, we will follow the alternate path, and show that the Lifshitz 
spacetime is a vacuum solution of minimal HL gravity, with no additional 
matter.   The preferred foliation of the Lifshitz spacetime, required for its 
embedding into HL gravity, is simply the foliation by leaves of constant 
$t$.  We will often split the bulk coordinates $y^\mu$ into time $t$ plus 
$D+1$ spatial coordinates $y^a=(x^i,r)$, $a=1,\ldots D$, and write the 
spacetime metric $G_{\mu\nu}$ in the Hamiltonian decomposition, 
$$
G_{\mu\nu}dy^\mu dy^\nu=-N^2dt^2+g_{ab}(dy^a+N^adt)(dy^b+N^bdt).
$$
Thus, $g_{ab}$ is the metric on the spatial bulk leaves of fixed $t$, $N_a$ 
is the shift vector and $N$ the lapse function.  Since the Lifshitz geometry 
(\ref{lifmet}) requires $N$ with a spatial dependence, we work in the 
nonprojectable version of HL gravity, with $N$ a full-fledged 
spacetime-dependent field.  Gauge symmetries are the foliation-preserving 
diffeomorphisms of spacetime.  

HL gravity may enjoy better short-distance properties than Einstein 
gravity (if it is dominated at high energies by its own $z>1$ scaling), 
but here we will follow the ``bottom-up'' strategy common in relativistic 
holography, and work only in the low-energy bulk gravity approximation.  
This is equivalent to the large $N$ limit in the dual field theory.  In this 
low-energy limit, HL gravity is dominated by the most relevant terms 
compatible with the gauge symmetries, and its effective action is 
\be
\label{hla}
S=\frac{1}{2\kappa^2}\int dt\,d^Dx\,dr\sqrt{g}N\left\{\vphantom{\frac{1}{2}}
K_{ab}K^{ab}-\lambda K^2+\beta(R-2\Lambda)+\frac{\alpha^2}{2}
\frac{\nabla_aN\nabla^aN}{N^2}\right\}.
\ee
(Here $K_{ab}=\frac{1}{2N}(\p_t g_{ab}-\nabla_a N_b-\nabla_bN_a)$ is the 
extrinsic curvature of the foliation, $R$ the scalar curvature of $g_{ab}$, and 
$K=g^{ab}K_{ab}$.)  The novelty compared to pure Einstein gravity is in the 
three couplings $\beta$, $\lambda$ and $\alpha$, which in Einstein gravity are 
fixed to $\lambda=\beta=1$ and $\alpha=0$.  Note that turning on the $\alpha$ 
coupling is important for the consistency of nonprojectable HL gravity 
\cite{bps2,grx}:  Taking the naive $\alpha\to 0$ limit in (\ref{hla}) would 
lead to a non-closure of the constraint algebra.  

When $\Lambda=0$, the flat spacetime $\R^{D+2}$ is a solution of (\ref{hla}).  
The propagating graviton modes consist of the transverse-traceless 
tensor polarizations with dispersion relation $\omega^2=\beta k^2$ (here 
$k\equiv\sqrt{k_ak_a}$ is the magnitude of the spatial momentum), plus an  
extra scalar graviton polarization, with dispersion
\be
\label{dispsc}
\omega^2=\frac{\beta(1-\lambda)}{\left[1-(D+1)\lambda\right]}\left[1+
D\left(\frac{2\beta}{\alpha^2}-1\right)\right]k^2.
\ee
The requirement of stability and perturbative unitarity around flat 
spacetime constrains the couplings to be in the range $\beta>0$, 
\be
\alpha^2\leq \frac{2\beta D}{D-1},
\ee
and 
\be
\label{lambrf}
\lambda\geq 1\quad {\rm or}\quad \lambda\leq 1/(D+1).
\ee

Turning on the cosmological constant 
$\Lambda<0$, we find that the Lifshitz geometry (\ref{lifmet}) is 
a vacuum solution of HL gravity with low-energy effective action (\ref{hla}), 
if
\be
\Lambda=-\frac{(D+z-1)(D+z)}{2}
\ee
and
\be
\alpha^2=\frac{2\beta(z-1)}{z}.
\ee
This simple interpretation of Lifshitz spacetimes as vacuum solutions of HL 
gravity suggests that the latter is the natural minimal holographic model of 
holographic duality for Lifshitz-type field theories.  

Further evidence for the universality of this minimal model of Lifshitz 
holography comes from the analysis of anisotropic Weyl anomalies 
in holographic renormalization, initiated in \cite{lgh} (and also later in 
\cite{jan}).  Just like their relativistic counterparts, anisotropic Weyl 
anomalies contain a lot of universal information about the system, and 
serve as useful probes of the duality.  Consider again some Lifshitz-type QFT 
on $\R^{D+1}$ with coordinates $(t,x^i)$ and a general background 
metric 
$$
ds^2=-\hat N^2dt^2+\hat g_{ij}(dx^i+\hat N^idt)(dx^j+\hat N^jdt).
$$
It is useful to think of this theory as residing at the spacetime boundary 
$r\to\infty$ of the $D+2$ dimensional asymptotically Lifshitz spacetime, with 
$\hat g_{ij}$, $\hat N_i$ and $\hat N$ being the components of the 
appropriately defined (anisotropic conformal class of the) boundary metric 
\cite{lgh}.  Generally, we will put hats on quantities defined at the 
boundary, to distinguish them from their bulk counterparts.    
Define anisotropic Weyl transformations, generated by a spacetime dependent 
$\sigma(t,x^i)$:
\be
\label{aweyl}
\delta\hat g_{ij}=2\sigma\hat g_{ij},\quad\delta\hat N_i=2\sigma\hat N_i,\quad 
\delta \hat N= z\sigma\hat N.
\ee
These represent a local generalization of the rigid scaling transformations 
(\ref{scaling}).  A QFT which is classically invariant under (\ref{aweyl}) 
can develop an anisotropic Weyl anomaly at the quantum level, with the 
effective action transforming as (see Appendix~C of \cite{lgh} for details):
$$
\delta S_{\rm eff}[\hat g_{ij},\hat N_j,\hat N]=\int dt\,d^Dx\sqrt{\hat g}\hat 
N\,\sigma(t,x^i)\CA(t,x^i).
$$
The independent terms that can appear in $\CA(t,x^i)$ are local functionals of 
the metric, invariant under foliation preserving diffeomorphisms.  They 
can be classified by solving a cohomological problem 
\cite{lgh} designed to automatically incorporate the Wess-Zumino consistency 
conditions on the anomaly.  However, the multiplicative coefficients with 
which these terms contribute to the anomaly (and which we will refer to as 
``central charges'') must be calculated for a given theory on a case-by-case 
basis.  Perhaps the simplest nontrivial case is $D=2$ and $z=2$.  In this 
case, the anomaly is \cite{lgh}
\be
\label{anomaly}
\CA=c_K\!\!\left(\hat K_{ij}\hat K^{ij}-\frac{1}{2}\hat K^2\right)
+c_V\!\!\left(\hat R-\frac{\hat\nabla^i\hat N\hat\nabla_i\hat N}{\hat N^2}
+\frac{\hat\Delta \hat N}{\hat N}\right)^2,
\ee
with two independent central charges, $c_K$ and $c_V$.  (Here $\hat K_{ij}$, 
$\hat\nabla_i$ and $\hat R$ are the extrinsic curvature, connection 
and the scalar curvature constructed from $\hat g_{ij}$).  As noted in 
\cite{lgh}, $\CA$ in (\ref{anomaly}) takes the form of the Lagrangian for 
$z=2$ conformal HL gravity \cite{mqc,lif} in $2+1$ dimensions.  Moreover, 
while the first term in (\ref{anomaly}) satisfies the so-called ``detailed 
balance condition,'' ({\it i.e.}, it is related to the square of the variation 
of another local functional $W$, see \cite{mqc,lif}), the other does not.  

For QFTs with holographic gravity duals, we can calculate the anomaly by 
performing holographic renormalization of the bulk theory 
\cite{mans,perfinn,dbvv} (see \cite{revsk,revdb,revjp} for reviews).  The 
relativistic Weyl anomaly was calculated this way in \cite{mans}.  
Holographic renormalization for the relativistic bulk theory (\ref{rela}) in 
asymptotically Lifshitz spacetimes was developed and applied to the 
anisotropic Weyl anomalies in \cite{lgh}, following the earlier work of 
\cite{omid,ross,mann,deboer}.  This procedure relies substantially on the 
notion of anisotropic conformal infinity developed in \cite{aci}.  In the 
low-energy gravity approximation, $S_{\rm eff}[\hat g_{ij},\hat N_i,\hat N]$ 
is calculated by evaluating the on-shell gravity action with the appropriate 
fall-off conditions on the metric field as $r\to\infty$.  This on-shell 
action is divergent due to inifinite volume, and needs to be renormalized.  
We regulate it by cutting $r$ off at $r_\epsilon=1/\epsilon$, and expand the 
on-shell action asymptotically to reveal the structure of its divergences.  
For the special case of $D=z$, this expansion gives \cite{lgh} (modulo terms 
that vanish as $\epsilon\to 0)$:
$$
\int dt\,d^Dx\sqrt{\hat g}\hat N\left\{\sum_{n=0}^{D-1}\frac{\CL^{(2n)}}{
\epsilon^{2(D-n)}}-\tilde \CL^{(2D)}\log\epsilon+\CL^{(2D)}\right\}.
$$
(The $\log\epsilon$ term is present only for special values of $D$ and $z$ 
\cite{lgh}, including the case $D=z=2$ of interest here.)  
The divergent terms are then cancelled by local counterterms, and 
$S_{\rm eff}=\int dt\,d^Dx \sqrt{\hat g}\hat N\CL^{(2D)}$.  

To calculate these divergent terms, we use the Hamiltonian form of holographic 
renormalization, as 
developed for relativistic AdS/CFT in \cite{papask,papaske} and extended 
to the asymptotically Lifshitz case in \cite{lgh} following \cite{ross}.  
In this formulation, the on-shell bulk action is determined as a functional 
of the boundary metric because it satisfies the Hamilton-Jacobi (HJ) equation 
for the radial evolution along $r$.  The operator of radial evolution 
$\delta_\CD$ is given by the generator of anisotropic dilatations 
on the boundary, and the HJ equation yields a recursive relation between the 
divergent terms $\CL^{(m)}$ of adjacent scaling dimensions $m$.  One of these 
recursive relations implies \cite{lgh}
\be
\delta_\CD\CL^{(2D)}=-2D\CL^{(2D)}+\tilde\CL^{(2D)}.
\ee
Interpreted from the boundary point of view, this means that when 
$\tilde\CL^{(2D)}\neq 0$, $S_{\rm eff}$ scales anomalously under the 
$z=D$ anisotropic Weyl transformations, and $\tilde\CL^{(2D)}$ is the 
anisotropic Weyl anomaly.

For the special case of the relativistic model (\ref{rela}) with $D=z=2$, 
the divergent terms were calculated in \cite{lgh}, where we obtained for 
the anisotropic Weyl anomaly
\be
\CA\equiv\tilde\CL^{(4)}=\frac{1}{16\pi G_{\rm N}}\left(\hat K_{ij}\hat K^{ij}
-\frac{1}{2}\hat K^2\right).
\ee
Thus, the anomaly in this relativistic model turns out to have $c_V=0$, or 
in other words, satisfies the detailed balance condition.  Why is it so?  
The conclusive answer was found in \cite{lgh}:  The relation 
implying that the anomaly in the relativistic model (\ref{rela}) should 
satisfy detailed balance follows from the holographic recursion relation 
between the divergent terms $\CL^{(2)}$ and $\tilde\CL^{(4)}$, with $\CL^{(2)}$ 
effectively playing the role of the local functional $W$ (see \cite{lgh} 
for details).  

Looking for holographic duals of more general QFTs with both central charges 
independently nonzero is an interesting challenge.  Before we embark on this  
pursuit, we should first check that QFTs whose central charges 
$c_K$ and $c_V$ are both nonzero indeed exist.  Examples of strongly coupled 
Lifshitz field theories are very scarce to say the least, but our point can 
be made by considering the theory of the free $z=2$ Lifshitz scalar $\Phi$.  
When $\Phi$ is minimally coupled to background HL gravity, 
$$
S_\Phi=\int dt\,d^2x\sqrt{\hat g}\left\{\frac{1}{\hat N}\left(\p_t\Phi-
\hat N^i\hat\nabla_i\Phi\right)^2-\hat N\left(\hat\Delta\Phi\right)^2\right\},
$$
this theory is classically invariant under (\ref{aweyl}) (with 
$\delta\Phi=0$), but develops an anisotropic Weyl anomaly at the quantum 
level.  This anomaly was calculated in \cite{jan}, and it turns out to have 
$c_V=0$.  One could perhaps speculate that $c_V=0$ might be a universal 
property of all consistent QFTs, hence eliminating the need for finding 
gravity duals with $c_V\neq 0$.  A simple counterexample comes from 
coupling $\Phi$ to background gravity non-minimally, adding 
$$
{}-e^2\int dt\,d^Dx\sqrt{\hat g}\hat N\left\{\hat R-\frac{\nabla^i\hat N\nabla_i
\hat N}{\hat N^2}+\frac{\Delta\hat N}{\hat N}\right\}^2\!\Phi^2
$$
to $S_\Phi$.  Even with this non-minimal coupling, this theory stays 
classically invariant under the anisotropic Weyl transformations 
(again with $\delta\Phi=0$), and develops a quantum anomaly.  We calculated 
this anisotropic Weyl anomaly using the $\zeta$-function regularization, and 
found $c_K=1/(32\pi)$ and $c_V=-e^2/(8\pi)$.  

Having demonstrated the existence of QFTs with $c_V\neq 0$, we can now ask 
how to reproduce this second central charge in a holographic gravity dual.  
One could look for relativistic models more complicated than (\ref{rela}).  
Instead, we will show that minimal HL gravity already accounts for both of 
the independent central charges $c_K$ and $c_V$ in the anisotropic Weyl 
anomaly.  In order to show that, we have performed holographic renormalization 
of Lifshitz spacetimes in pure HL gravity.  Since the technicalities are 
quite involved (as they were in the relativistic model (\ref{rela}) studied 
in \cite{lgh}), here we only present our main results; all technical details 
will appear in \cite{awl}.  

We find modified recursion relations for the divergent terms in the regulated 
action.  In the special case $D=z=2$, we solved these recursion relations 
and found that the logarithmic term $\tilde\CL^{(4)}$ is equal to
$$
\tilde\CL^{(4)}=\frac{1}{2\kappa^2}\left(\hat K_{ij}\hat K^{ij}
-\frac{1}{2}\hat K^2\right)
+\frac{\beta}{48\kappa^2}\left(\hat R-\frac{\hat\nabla^i\hat N\hat\nabla_i
\hat N}{\hat N^2}+\frac{\hat\Delta \hat N}{\hat N}\right)^2.
$$
This is the anisotropic Weyl anomaly in our minimal model of Lifshitz 
holography with vacuum HL gravity.  It is indeed of the most general form, 
with the two independent central charges given in terms of two low-energy 
couplings in minimal HL gravity: $c_K=1/(2\kappa^2)$ and 
$c_V=\beta/(48\kappa^2)$.   

The remaining coupling $\lambda$ does not appear in the anomaly, but it still 
plays an important physical role.  
Just as in the 
case of flat spacetime, $\lambda$ enters into the dispersion relation of 
the extra polarization of the graviton in the bulk.  For example in the 
radial gauge $g_{ri}=0$, $g_{rr}=1/r^2$, the extra graviton mode $\phi$ is 
found as the linearized fluctuation of the radial component of the shift 
vector, $N_r=\phi/r$.  Returning to the case of general $D$ and $z$, the 
linearized equations of motion imply the asymptotic 
behavior near infinity $\phi(r)\sim r^{D_{\pm}}$, with 
\be
D_{\pm}=\frac{1}{2}\left\{z-D\pm\sqrt{(z+D)^2
+\frac{4D(1-z)}{1-\lambda}}\right\}.
\ee
Standard rules of holographic duality will map $D_\pm$ to the scaling 
dimensions $\Delta_\pm$ of the operator dual to the extra graviton.  
Unitarity of the dual field theory requires that the scaling dimensions be 
real, implying (for $z>1$)
\be
\label{ubd}
\lambda\geq 1\quad{\rm or}\quad\lambda\leq\lambda_U\equiv
\frac{(D-z)^2+4D}{(D+z)^2}.
\ee
This unitarity bound represents an intriguing analog of the 
Breitenlohner-Freedman bound familiar from relativistic holography:   
In HL gravity, the unitarity bound (\ref{ubd}) allows the coupling $\lambda$ 
to dip into the region between $1/(D+1)$ and 1, which according to 
(\ref{lambrf}) would be forbidden around flat spacetime.  In the 
particularly interesting case of $D=z$, we get $\lambda_U=1/D$, which opens 
up the previously forbidden regime $1/(D+1)\leq\lambda\leq 1/D$.  

Now that we have seen that HL gravity provides candidate holographic duals 
for QFTs with anisotropic Lifshitz scaling, is it possible to apply HL 
gravity also to QFTs with isotropic $z=1$ scaling?  Interestingly, the limit 
$z\to 1$ corresponds to $\alpha\to 0$, the ``unhealthy reduction'' 
\cite{grx} of nonprojectable HL gravity, and may therefore be difficult to 
make sense of.  This is perhaps to be expected:  
$z=1$ QFTs with such gravity duals would likely exhibit isotropic 
dilatation symmetry without full relativistic conformal symmetry, a phenomenon 
whose examples are few and far between.  Further study of our holographic 
duality in the $\alpha\to 0$ limit may shed new light on this rare 
class of QFTs.  

Finally, throughout this paper we have used the effective low-energy limit of 
HL gravity, dominated by the terms of the lowest dimension in the action.  
We have been agnostic about how the model is completed at high energies.  
This completion may come from additional degrees of freedom, perhaps via an 
embedding into string theory; or it can be via a self-completion of HL 
gravity, due to highly anisotropic scaling at short distances.  This latter 
possibility would be particularly interesting, as it could open a new door 
away from the large $N$ limit and small bulk curvature.  As this paper was 
being finalized, complementary results about 
another form of nonrelativistic holography with HL gravity were presented in 
\cite{jk1,jk2}.  Our results, and those of \cite{jk1,jk2}, thus provide 
further evidence for the picture proposed originally in \cite{lgh}, that the 
natural arena for nonrelativistic holography is nonrelativistic HL gravity.  
It remains to be seen whether -- as suggested in \cite{lgh} -- the 
nonrelativistic field theories whose holographic duals happen to be 
relativistic indeed represent only a minority among all theories with gravity 
duals.

\acknowledgments

We wish to thank Kevin Grosvenor, Omid Saremi and Kostas Skenderis for useful 
discussions.  
The results presented in this paper were reported by one of the authors (PH) 
at the ``32nd Winter School on Geometry and Physics,'' Srn\'\i, \v Sumava, 
Czech Republic (January 2012), the workshops ``Exploring the Quantum 
Spacetime,'' Bad Honnef, Germany (March 2012) and ``Quantum Gravity in 
Paris,'' Paris, France (March 2012), the ``6th International School on Field 
Theory and Gravitation,'' Petr\'{o}polis, Brazil (April 2012), ``The Conformal 
Universe'' workshop, Perimeter Institute, Waterloo, Canada (May 2012), 
in a plenary talk at the ``13th Marcel Grossmann Meeting MG13,'' Stockholm, 
Sweden (July 2012), and at the ``2nd BCTP Summit at Lake Tahoe,'' Glenbrook, 
Nevada (September 2012); and by one of us (CMT) at the University of 
Washington (April 2012).  PH and CMT wish to thank the organizers and hosts 
of these events for their wonderful hospitality.  
This work has been supported in part by NSF Grants PHY-0855653 
and PHY-1214644, DOE Grant DE-AC02-05CH11231, and BCTP.  

\bibliographystyle{JHEP}
\bibliography{lhj}
\end{document}